\documentclass[prd,preprintnumbers]{revtex4}
\usepackage{amsmath}
\usepackage{amsfonts}
\usepackage{graphicx}

\begin{document}
\tolerance=5000
\def\pp{{\, \mid \hskip -1.5mm =}}
\def\cL{{\cal L}}
\def\be{\begin{equation}}
\def\ee{\end{equation}}
\def\bea{\begin{eqnarray}}
\def\eea{\end{eqnarray}}
\def\tr{{\rm tr}\, }
\def\nn{\nonumber \\}
\def\e{{\rm e}}

\preprint{}


\title{A note on dark energy induced by D-brane motion}

\author{R. Chingangbam}
\affiliation{Centre for Theoretical Physics, Jamia Millia Islamia,
New Delhi-110025, India} \email{ronidchi@gmail.com}

\author{A. Deshamukhya}
\affiliation{Department of Physics,Assam University,
Silchar-788011,India} \email{atri.deshamukhya@gmail.com}

\begin{abstract}
 In this note we study the possibility of obtaining dark energy solution in a D-brane scenario in a
warped background that includes brane-position dependent corrections
for the non-perturbative superpotential. The volume modulus is
stabilized at instantaneous minima of the potential. Though the
model can account for the existence of dark energy within present
observational bound -- fine-tuning of the model parameters becomes
unavoidable. Moreover, the model does not posses a tracker solution.

\end{abstract}

\maketitle

\section{Introduction}

Study of high red-shifted supernovae and other cosmological
observations \cite{ARIES} clearly indicate that the expansion of the
universe is accelerating rather than slowing down as an expected
result of gravitational pull. Within the framework of the standard
cosmological model, this implies that 70 percent of the universe is
composed of a new, mysterious {\it dark energy}\cite{
SS,PADDY,EMS} which counters the attractive force of gravity
unlike any known form of matter or energy  .

Known to be very homogeneous, not very dense and non-interacting
other than gravitationally, the exact nature of dark energy is yet a
matter of speculation. Dark energy appears to be the dominant
component of the physical Universe, but there is no persuasive
theoretical explanation for its existence or magnitude as on date.

The simplest possible explanation for dark energy is that it's the
'cost of having space'. In other words, the Universe is permeated by
an energy density, constant in time and uniform in space. General arguments from the scale of particle interactions,
however, suggest that if $\Lambda$ is not zero, , it would be
expected to be $10^{120}$  times larger than what is observed.

 An important step towards a realistic cosmological model based on
string theory came from the realization that background fluxes can
stabilize most of the moduli of string theory. It was shown in
\cite{GKP} that fluxes in warped compactifications, using a
Klebanov-Strassler (KS) throat \cite{KS} can stabilize the dilaton
and complex structure moduli of type IIB string theory compactified
on an orientifold of a Calabi-Yau threefold. Infact, it was shown in
\cite{KKLT} that all the closed string moduli can be stabilized by a
combination of fluxes and non-perturbative effects. The
non-perturbative effects are mainly responsible for stabilizing the
Kahler muduli and they arise either from an Euclidean D3-brane
wrapping a four cycle or from gauge dynamics of a stack of $n$
D7-branes wrapping supersymmetrically a four cycle in the warped
throat.

The D-brane physics  finds interesting applications to early
universe and late time cosmology. The efforts of constructing
inflationary models exploiting D-brane dynamics is still under
active consideration. Nevertheless it is important to remember that
a viable inflation apart from meeting observational constraints
also should be followed by a successful reheating.This severely constraints building inflationary models. Based on this it is not easy to build such a
model in the frame work of D-brane dynamics\cite{BDKM, PST, BDKMS}.
However, it is simpler to build a dark energy model which is free from such constraints.

In this letter, we address the possibility of having a viable dark
energy model from the effective scalar field potential $V (\phi)$
obtained in Ref.\cite{BDKMS} after the volume mudulus is fixed at
the instantaneous minimum.

\section{Effective  field description}

 The inflaton potential can
be obtained by performing string theoretic computations involving
the details of the compactification scheme. In this setup, the
inflation is realized by the motion of a D3-brane towards a distant
static anti-D3-brane, placed at the tip of the throat. The position
of the moving brane in the compactification manifold is identified
with the inflaton field. To be more precise, the location of the
mobile brane can be labeled with five angular coordinates and one
radial coordinate, $r$. The canonical inflaton field $\phi$ can be
expressed by a constant rescaling of this radial coordinate. The
stability analysis for the trajectories in the angular directions
has to be performed consistent with the embedding of D7-branes.
However, it was shown in \cite{BDKMMM}, with the embedding of the
D7-branes as given in \cite{Kup} and requiring at least one of the
four-cycles carrying nonperturbative effects to descend down a
finite distance into the warped throat, that the presence of a
D3-brane gives rise to a perturbation to the warp factor affecting a
correction to the warped four cycle volume. Moreover, this
correction depends on the position of the D3-brane and thus the
superpotential for the nonperturbative effect gets corrected by an
overall position dependent factor. The total potential that the
inflaton field experiences is the sum of the potential (F-term)
coming from the superpotential as mentioned above and the usual
D-term potential coming from the interaction between the D3-brane
and the anti-D3 brane. Note that the presence of the anti-D3-brane
breaks the supersymmetry of the system and the scale of the
supersymmetry breaking is given by $D_0~=~2 T_3 h_0^{-1}$ where
$h_0$ is the warp factor at the bottom of the throat. It depends on
the fluxes and approximately can be written as $h_0 \sim e^{8\pi
K/(3 g_s M)}$ where $g_s$ is the string coupling constant and
$M$-units of $F_3$ flux and $-K$-units of $H_3$ flux are turned on,
respectively, through the A-cycle and the dual B-cycle at the tip of
the throat. The issue of volume modulus stabilization, thus needed
to be re-analysed and has been carried out in the recent work of
\cite{BDKMS}, \cite{BDKM}, \cite{KP} and \cite{PST} and in this
changed scenario, the viability of inflation has been readdressed
and it has been observed that delicate fine tunings of the
parameters are necessary to chase the inflation.

The models in Refs.[\cite{BDKMS},\cite{BDKM}]is described by a
two-field potential $V(\sigma, \phi)$ in terms of the inflaton
$\phi$ and the volume modulus $\sigma$. We do not repeat here the derivation of the potential for which one can consult the Ref.[\cite{PST}].If the mass of the modulus is much larger than
Hubble rate, the field $\sigma$ evolves along the insantaneous
minima determined by the condition $V_{,\phi}=0$ which determines
$\sigma$ (equal to $\sigma_\star$) in terms of the field $\phi$. The
two field potential $V(\phi, \sigma)$, thus can be reduced to an
effective one field potential when the field $\sigma$ continues to
remain in its instantaneous minimum $\sigma_\star(\phi)$ which
evolves slowly. The effective single field potential then acquires
the following form,

\begin{eqnarray}
 V(\phi) &=& \frac{ a }{3} \frac{\exp (-2 w_0 \sigma_\star(\phi))}
{ U(\phi,\sigma_\star(\phi))^2}  g(\phi)^{2/n} \Biggl[
2 w_0 \sigma_\star(\phi) +6- 6 \exp (w_0\sigma_\star(\phi) ) 
|W_0| \frac{1}{g(\phi)^{1/n}}  \nonumber \\
&&  + \frac{3c}{n}  \frac{\phi}{\phi_\mu} \frac{1}{g(\phi)^2} - \frac{3}{n} \left( \frac{\phi}{\phi_\mu} \right)^{3/2} \frac{1}{g(\phi)^2}- \frac{3}{n} \left( \frac{\phi}{\phi_\mu} \right)^3 \frac{1}{g(\phi)^2} \Biggr] + \frac{D(\phi)}{U(\phi,\sigma_\star(\phi))^2} \label{equ:V}
\end{eqnarray}

where the functions $\sigma_\star(\phi)$, $g(\phi)$, $U(\phi)$ and
$D(\phi)$ are given by,
\bea
&&\sigma_\star(\phi) \approx \left[1 +
\frac{1}{n} \frac{1}{w_F}\left(1-\frac{1}{2w_F} \right)
\left( \frac{\phi}{\phi_\mu} \right)^{3/2}\right]\, ,\\
&&g(\phi)=1+\left(\phi/\phi_\mu\right)^{3/2}\, , \\
&& U(\phi)=\frac{2\sigma_\star w_0}{a}-\frac{w_0}{3a}(\phi/\phi_\mu)^2\phi_\mu^2 \, , \\
&&D(\phi)=D_0\left(1-\frac{27D_0}{64\pi^2(\phi/\phi_\mu)^4\phi_\mu^4}\right)
\label{equ:sigma}
\end{eqnarray}

with the notations

\begin{equation} \label{equ:c32}
c\equiv \frac{9}{4nw_0\phi_\mu^2},~~w_F\equiv a \sigma_F,~~w_0\equiv
a\sigma_0,~~a=\frac{2\pi}{n}
\end{equation}
Here $n$ designates the number of $D7$ branes. The constants $W_0$,
$w_F$ and $w_0$ are constrained by the following relations
\begin{eqnarray}
&&3\frac{|W_0|}{|A_0|}e^{a\sigma_F}=2a\sigma_F+3 \,, \\
&&3\frac{|W_0|}{|A_0|}e^{w_0}=2w_0+3+s,~~~~1<s<3
\end{eqnarray}

In what follows, we shall try to use the field $\phi$ with effective
potential (\ref{equ:V}) as a quintessence field.

\begin{figure}
\label{fig1}
\begin{center}
\includegraphics[width=9cm,height=7cm,angle=0]{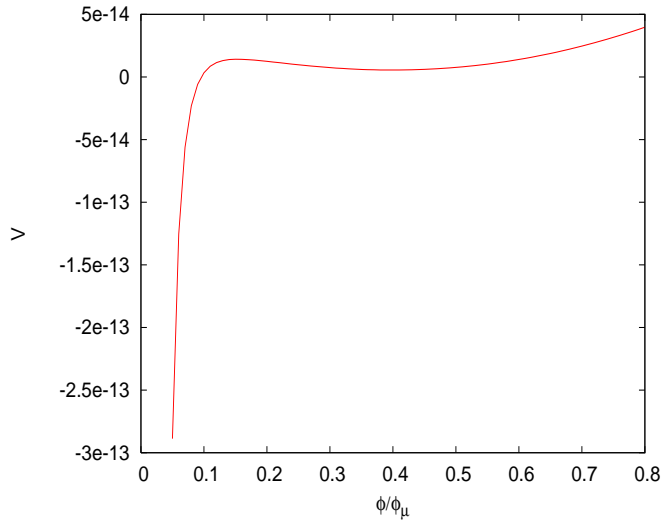}
\end{center}
 \caption{Plot of the potential, $V$ under single field approximation for the model parameters $n=8$, $\omega_0=10.1$, $\omega_F=9.9951$, $W_0=3.496 \times 10^{-4}$, $D_0=1.099\times 10^{-8}$ and $\phi_{\mu}=0.25$ }.
\end{figure}

\section{Dark energy}
Before we get to numerics, we would like to emphasize the
interesting features of the potential (\ref{equ:V}[Fig:1]). As
pointed out in Ref.\cite{BDKM}, one can choose the parameters in
the potential namely $\sigma_F$, $n$, $c$ and $D_0$ such that the
potential looks flat for $\phi$ smaller than the point of inflection
as $\phi$ moves toward zero ( $A_0$ gives the over all scale in the
potential and can be chosen as equal one) . For $\phi$ approaching
zero, the $D(\phi)$ term in the potential plays the deciding role as
$V(\phi) \sim D(\phi)\sim -(\phi/\phi_\mu)^{-4}$ making $V(\phi)$
steep and large negative near the origin. $V(\phi)$ is positive and
steep for $\phi \to \phi_\mu$. It further important to note that
$D_0$ controls the height of the flat part of the potential which
should mimic the cosmological constant like behavior; tuning $D_0$
can lead to the present day scale of dark energy. These are
essential features that a model should exhibit in order to account
for the late time acceleration.

By fixing the constants and parameters as,
\begin{eqnarray}
&& w_F=9.9956,\nonumber,~~w_0=10.1, \nonumber~~ n=8\\
 && a=\frac{2
\pi}{n}, ~~ \phi_\mu=0.25,~~
 W_0=3.496 \times 10^{-4} \nonumber
\end{eqnarray}

the effective single field potential as a function of $\phi /
\phi_{\mu}$ and $D_0$ can be expressed as a Taylor series expansion
around any particular value of $\phi /\phi_{\mu}$ in the flat region
of the potential. The potential is flat around $\phi / \phi_{\mu} =
0.4$.  The flat part of potential can be expressed as,

\begin{eqnarray}
&&{V}\left(\frac{\phi}{\phi_{\mu}}, D_0\right) \simeq
{-1.6544\times10^{-11}}+ 0.0015 D_0-0.6444 {D_0}^2
\end{eqnarray}

Since the value of $D_0$ is expected to be of the order of $10^{-8}$
from string theory point of view, one can observe that by fine
tuning of $D_0$, it is possible to realize a cosmological constant
within present observational bound. It corresponds to the case when
$ D_0 \simeq (1.6544 \times 10^{-11})/0.0015$. It is further
desirable that the field energy density should mimic the background
being subdominant soon after the commencement of the radiative
regime and it should take over the background energy density at late
times thereby alleviating the fine tuning and coincidence problems.
In what follows we shall investigate the cosmological viability of
D-brane scenario in context with dark energy.

The Friedmann equation and field equation
 \bea
&&H^2=\frac{1}{3}\left(\dot{\phi}^2/2+V(\phi)+\rho_m\right),\\
&&\ddot{\phi}+3H\dot{\phi}+V'(\phi)=0 \eea can be cast as first
order equations \bea
&&\frac{dx}{dN}=y/H(x), \\
&&\frac{dy}{dN}=-3y-U_x(x)/H(x) \\
&& H(x)=\phi_\mu \sqrt{\frac{1}{3}\left(y^2/2+U(x)+\phi_\mu^{-2}
\rho_m\right)} \eea
 where $x=\phi/\phi_\mu$,
$y=\dot{\phi}/\phi_\mu$, $U=V/\phi_\mu^2$ $N=\ln a$ and
$\rho_m=\rho^m_0e^{-3N}$ is the matter energy density.

We numerically investigated the dynamics described by the single
field potential (\ref{equ:V}); our results are depicted in Figures 2
and 3. We show the evolution of field energy and background (matter)
energy densities versus $N$ along with the dimensionless density
parameter.

\begin{figure}[h]
\label{fig2}
\begin{center}
\includegraphics[width=8cm,height=6cm,angle=0]{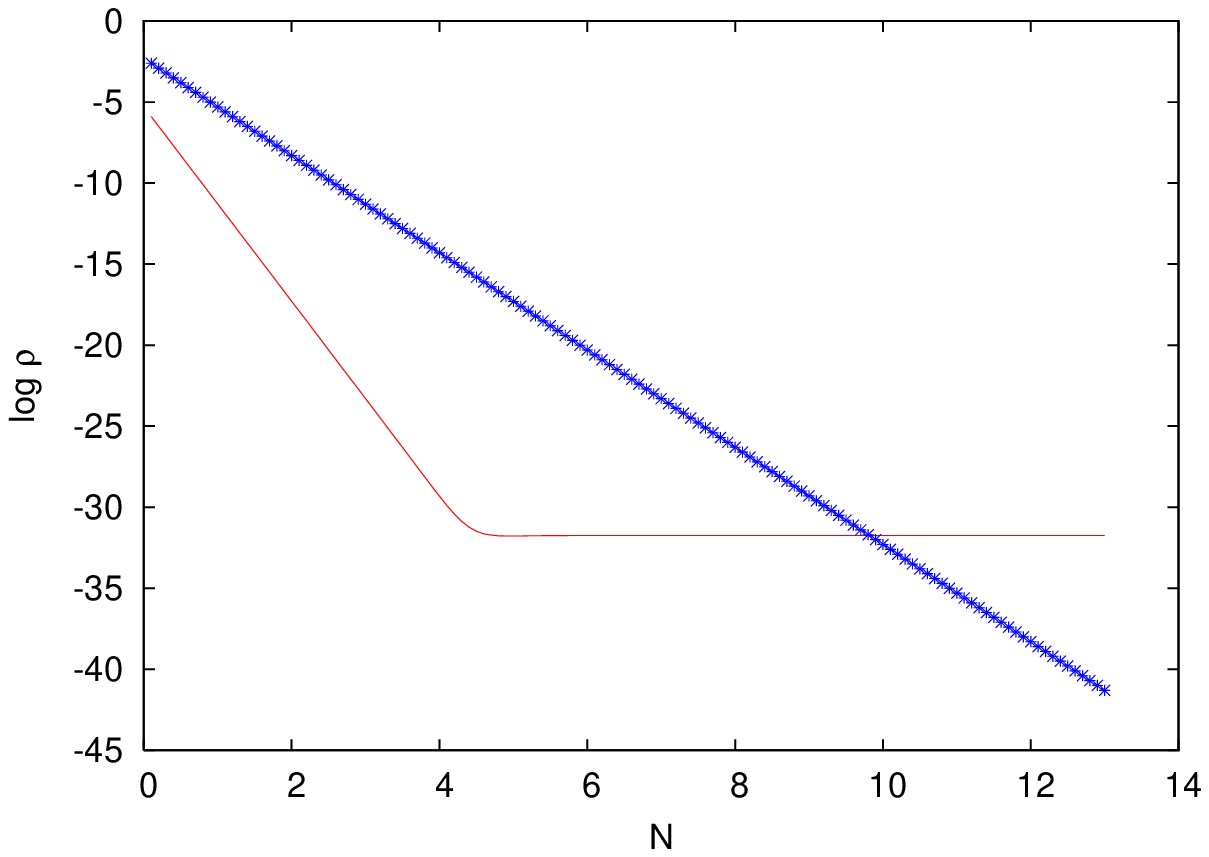}
\includegraphics[width=8cm,height=6cm,angle=0]{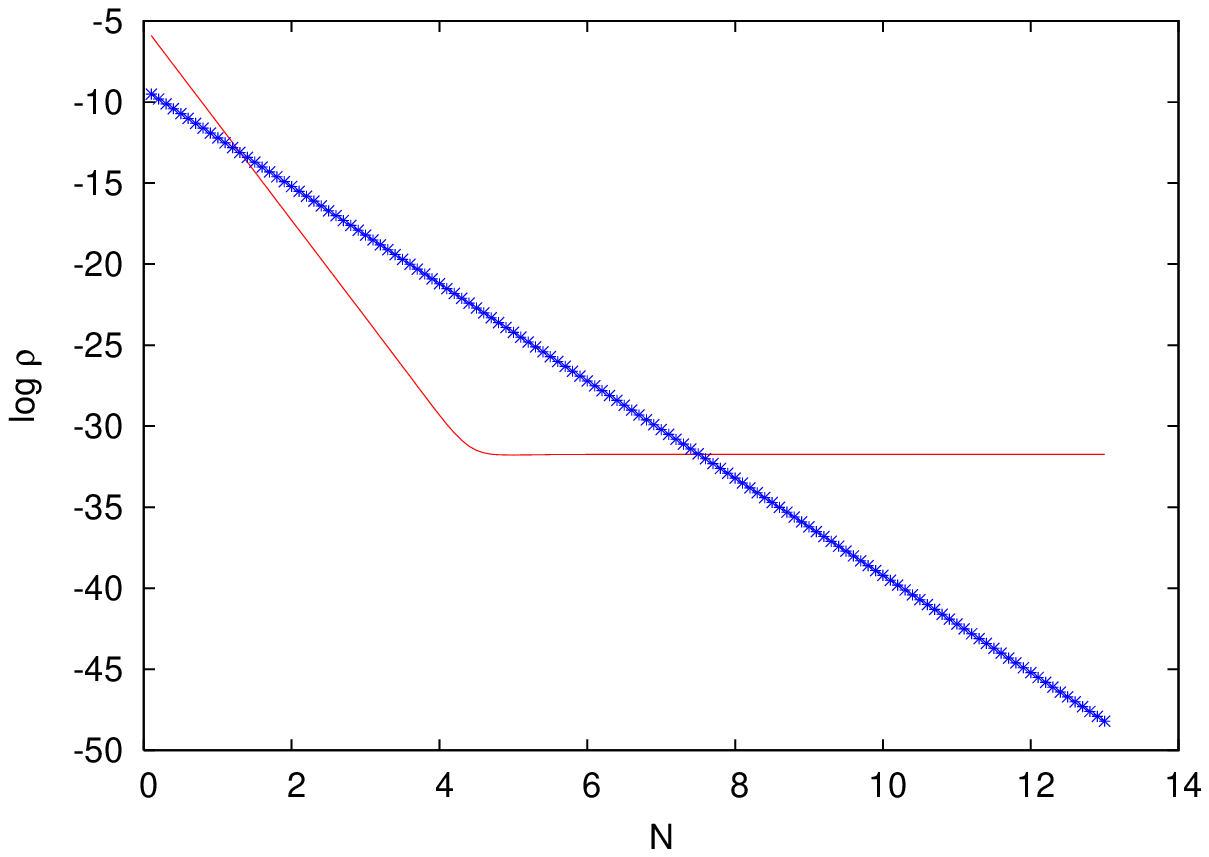}
\end{center}
 \caption{Plot of field energy and matter densities versus $N$ in case of overshoot(left) and undershoot(right)}.
\end{figure}
The right and left figures in the Fig. 2 display the cases of
undershoot and overshoot respectively. In both the cases the field
energy density continues scaling faster than the background energy
density till the field rolls along the steep part of the potential,
it then freezes mimicking the cosmological constant like behavior.
When the matter density becomes comparable to field energy density,
it begins evolving slowly and takes over the background to account
for the dark energy. It is clear that the model under consideration
does not possess tracker solution. For the tracker to exist, it is
necessary that the field potential remains steep close to the
exponential potential for most of the history of universe and
becomes shallow only at late times. Unlike the tracker solution, the
present scenario exhibits dependence on the initial conditions. Thus
apart from the tuning of the model parameters, the adjustment of the
initial conditions is necessary to obtain the observed accelerated
expansion.

To summarize, we have examined the possibility of late acceleration
using single field D-brane potential. This scenario was applied to
inflation earlier; it is perfectly legitimate to apply the same to
late time acceleration with little change of parameters. We have
shown that de-Sitter is a late time attractor of the model. The
present value of the cosmological constant can be obtained by fine
tuning the value of the constant $D_0$. The absence of a tracker
solution gives rise to additional dependency on the initial
conditions of the field $\phi$.

\begin{figure}[h]
\label{fig3}
\begin{center}
\includegraphics[width=8cm,height=6cm,angle=0]{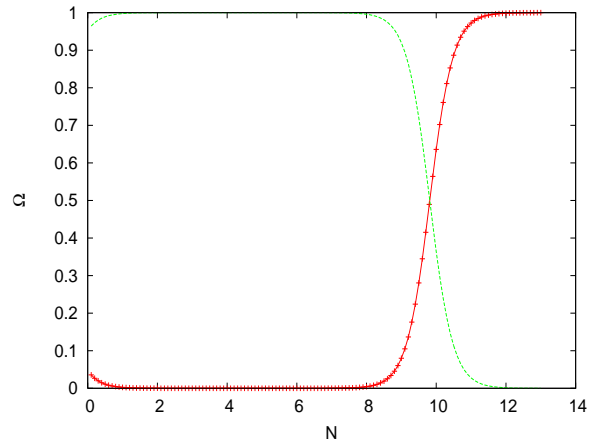}
\end{center}
 \caption{The evolution
 of dimensionless density parameter $\Omega$ with $N$}.
\end{figure}

\section{Acknowledgements}

 The authors would like to thank M Sami , CTP, Jamia Millia Islamia and S Panda, HRI, Allahabad
 for their valuable suggestions. AD thanks CTP,Jamia for supporting her visit to the centre during
 the period over which this work has been carried out.



\begin{thebibliography}{99}
\bibitem{ARIES} A Albrecht et al, astro-ph/0609591
\bibitem{SS} V Sahni and A Starobinsky, Int.J.Mod.Phys. D15 (2006) 2105-2132
\bibitem{PADDY} T Padmanabhan, astro-ph/0602117
\bibitem{EMS} E J Copeland, M Sami and Shinji Tsujikawa, Int.J.Mod.Phys. D15 (2006) 1753-1936
\bibitem{GKP} S B Giddings, S. Kachru and J. Polchinski, Phs.Rev D {\bf{66}}, 106006 (2002)
\bibitem{KS} I. R. Klebanov an M. J. Strassler, J. High Energy Physics. 08 (2000) 052
\bibitem{KKLT} S. Kachru, R. Kallosh, A. Linde and S. P. Trivedi, Phys Rev D {\bf{68}}, 046005(2003)
\bibitem{BDKMMM} D. Baumann , A. Dymarsky, I. R. Klebanov, J. Maldacena, L. McAlliter and A Murugan, J.High Energy Phys. {\bf{11}} (2006) 031
\bibitem{Kup} S Kuperstein, J. High Energy Phys
ics. 03 (005) 014
\bibitem{BDKMS} D. Baumann, A. Dymarsky, I. R. Klebanov, L. McAllister and P. J. Steinhardt, Phys. Rev Lett.  {\bf{99}}, 141061 (2007)
\bibitem{BDKM} D. Baumann, A. Dymarsky, I. R. Klebanov and L.
McAllister, arXiv: 070.0360.
\bibitem{KP} A. Krause and E. Pajer, arXiv:0705.468
\bibitem{PST} S. Panda, M. Sami and S Tsujikawa, Phys. Rev {\bf{D76}} 103512 (2007)


\end{thebibliography}
\end{document}